\documentclass[epj,final]{svjourpp}

\usepackage{graphicx}
\usepackage{amssymb}
\usepackage{amsmath}
\usepackage{bm}
\usepackage{multirow}
\usepackage{color}

\def\be{\begin{eqnarray}}
\def\ee{\end{eqnarray}}

\begin{document}

\title{Neutron drip line in the deformed relativistic Hartree-Bogoliubov theory in continuum: Oxygen to Calcium}
\author{Eun Jin In\inst{1}\and Youngman Kim\inst{2}\and Panagiota Papakonstantinou\inst{2}\thanks{\email{ppapakon@ibs.re.kr}} \and Seung-Woo Hong\inst{3}}
\institute{Department of Energy Science, Sungkyunkwan University, Suwon 16419, Korea 
\and Rare Isotope Science Project, Institute for Basic Science, Daejeon 34000, Korea
\and Department of Physics, Sungkyunkwan University, Suwon 16419, Korea 
} 
\authorrunning{Eun Jin In et al.}
\titlerunning{Neutron drip line in DRHBc: O to Ca}

\date{\today}

\abstract{The location of the neutron drip line, currently known for only the lightest elements, remains a fundamental question in nuclear physics. Its description is a challenge for microscopic nuclear energy density functionals, as it must take into account in a realistic way not only the nuclear potential, but also pairing correlations, deformation effects and coupling to the continuum.
The recently developed deformed relativistic Hartree-Bogoliubov theory in continuum (DRHBc) aims to provide a unified description of even-even nuclei throughout the nuclear chart. 
Here, the DRHBc with the successful density functional PC-PK1 is used to investigate whether and how deformation influences the prediction for the neutron drip-line location for even-even nuclei with $8 \leq Z \leq 20$, where many isotopes are predicted deformed. 
The results are compared with those
based on the spherical relativistic continuum Hartree-Bogoliubov (RCHB) theory and discussed in terms of shape evolution and the variational principle.
It is found that the Ne and Ar drip-line nuclei are different after the deformation effect is included. The direction of the change is not necessarily towards an extended drip line, but rather depends on the evolution of the degree of deformation towards the drip line. 
Deformation effects as well as pairing and continuum effects treated in a consistent way can affect critically the theoretical description of the neutron drip-line location.
}

\maketitle

\section{INTRODUCTION}

Nuclear masses are not only of great importance in nuclear physics but are also of interdisciplinary interest notably for research on fundamental interactions and astrophysics~\cite{Lunney2003}.
While ongoing experiments reach ever farther away from the valley of stability, the neutron drip line has only been mapped for the lightest elements
 and theory is essential for providing qualitative and quantitative predictions and extrapolations.
The theoretical research on the neutron drip line and associated exotic phenomena
has long been active~\cite{Meng2006,Meng2015,Zhou2016,Gupta2000,Matsuo2010,Hamamoto2012,Afanasjev2015,Wang2015,Neufcourt2019}.
The neutron-rich nuclei show different structures
and properties
compared to the stable nuclei:
As the number of neutrons increases,
the shape \textcolor{black}{and size of nuclei may change in ways not observed in stable nuclei, leading to new phenomena like giant haloes, soft and pygmy resonances, and new magic numbers, in addition to neutron skin and new regions of shape coexistence and isomerism} 
~\cite{Meng2015,Gupta2000,Savran2013,Otsuka2020,Tanihata1995,Meng1998,Egelhof2002,Tanihata2003,Stoitsov2003,Long2010,Thakur2019,Saxena2019}, 
with implications for astrophysical and nucleosynthesis studies.

In unstable, weakly-bound nuclei close to the neutron drip line,
continuum and pairing correlations play an important role~\cite{Meng2006,Meng2015,Zhou2016,Meng1998,Hansen1987,Kucharek1991,Meng1996,Meng2002}.
The neutron Fermi energy is very close to the continuum, so in order to study theoretically the neutron drip line, pairing correlations and continuum effects
must be taken into account to determine the neutron separation energy as precisely as possible.
Both pairing correlations and the treatment of continuum effects can influence the predictions for the drip line.

The Hartree-Fock-Bogoliubov (HFB) model has found many applications in the study of nuclear masses and the drip lines~\cite{Goriely2009sky,Goriely2009gog,Erler2012}. 
The relativistic Hartree-Bogoliubov approach (RHB) has been used in the study of haloes in spherical nuclei~\cite{Meng1996,Poschl1997,MengRHB1998}.
Recently,
the Relativistic Continuum Hartree-Bogoliubov (RCHB) theory~\cite{Meng1996,MengRHB1998} 
was used to construct a complete mass table and explore nucleon drip-lines~\cite{Xia2018}
by assuming spherical symmetry throughout. 
\textcolor{black}{The PC-PK1 relativistic energy density functional was used, 
which has been shown to provide a good description of infinite nuclear matter and finite nuclei, including the isospin dependence of the binding energy along isotopic and isotonic chains}~\cite{Zhao2010}.
The ground-state properties of nuclei
with proton number 8 to 120 were investigated.
With the effects of the continuum included,
there are totally 9035 nuclei predicted to be bound.
In that work,
it was found that
the coupling between the bound states and the continuum
due to the pairing correlations
plays an essential role in extending the nuclear landscape.

\textcolor{black}{Most known nuclei are deformed. 
In order to properly describe deformed exotic nuclei, 
the deformed relativistic Hartree-Bogoliubov theory in continuum (DRHBc) was developed, initially based on the meson-exchange density functional~\cite{Zhou2010,Li2012} and extended to the version with density-dependent meson-nucleon couplings~\cite{Chen2012} and to include the blocking effect~\cite{Li2012b}.  
Continuum effects are included by solving the deformed RHB equations in a Dirac Woods-Saxon basis~\cite{Zhou2003}. 
Recently, the DRHBc with point-coupling density functionals was developed~\cite{Zhang2020}. 
The aim is to provide a unified description of even-even nuclei throughout the nuclear chart by including all important effects, namely continuum, pairing, and deformation, presently restricted to axial symmetry. 
The DRHBc theory in its various iterations has been applied in the study of exotic Mg isotopes, predicting a decoupling between the core and the halo in $^{42,44}$Mg~\cite{Zhou2010,Li2012}, in resolving the puzzle concerning the radius and configuration of valence neutrons in $^{22}$C~\cite{Sun2018}, and a study of particles in the classically forbidden regions of Mg isotopes~\cite{Zhang2019}.  
}  

In the present investigation, we are interested in how the inclusion of deformation degrees of freedom can affect the predictions for the drip line. 
The theoretical question posed is whether the additional degrees of freedom represented by deformation and generally leading to more binding 
can lead to a sizable extension of the nuclear landscape. 
To this aim, we compare new results obtained with the DRHBc with the 
\textcolor{black}{RCHB predictions of Ref.~\cite{Xia2018} and of  
Ref.~\cite{Qu2013}, which focuses on the region from Oxygen to Titanium. 
In both the DRHBc and the RCHB calculations, the PC-PK1 functional is used.} 
We focus on even-even nuclei in  the nuclear-chart region from Oxygen to Calcium.

This paper is structured as follows.
In Sec.~\ref{Sec:Method} the DRHBc formalism \textcolor{black}{and numerical implementation are} presented briefly.
In Sec.~\ref{Sec:Results} we present and discuss our results.
First, we examine how the deformation parameter evolves along the isotopic chains. 
Second, we report how the drip-line predictions change with respect to RCHB predictions, i.e., when we include deformation.
Finally, we inspect the evolution of separation energies towards the drip line in Ne, Mg, and Ar isotopic chains in order to interpret and discuss the results.
We conclude in Sec.~\ref{Sec:Conclusions}.

\section{DRHBc theory \label{Sec:Method}}

\subsection{Formalism} 
To describe the finite nuclear system,
the starting point is presently a Lagrangian density
of the point-coupling model~\cite{MengB2016},
\be
{\cal L} &=& \bar{\psi} \left(i\gamma_\mu \partial^\mu - m \right)\psi\,
                                                                 \nonumber\\
&-&\frac{1}{2} \, \alpha_{S} \left(\bar{\psi}\psi \right) \left(\bar{\psi}\psi \right)\,
-\frac{1}{2} \, \alpha_{V} \left(\bar{\psi} \gamma_\mu \psi \right) \left(\bar{\psi} \gamma^\mu \psi \right)\,
 \nonumber \\
    &-& \frac{1}{2} \, \alpha_{TV} \left(\bar{\psi} \vec{\tau} \gamma_\mu \psi \right)
 \left(\bar{\psi} \vec{\tau} \gamma^\mu \psi \right)\,
                                                     \nonumber\\
&-&\frac{1}{3}\beta_S \left(\bar{\psi}\psi \right)^3
-\frac{1}{4}\gamma_S \left(\bar{\psi}\psi \right)^4
-\frac{1}{4} \, \gamma_{V} \left[\left(\bar{\psi} \gamma_\mu \psi \right)
\left(\bar{\psi} \gamma^\mu \psi \right)\right]^2\,
                                                  \nonumber\\
&-&\frac{1}{2} \delta_S \partial_\nu \left(\bar{\psi}\psi \right)
 \partial^\nu \left(\bar{\psi}\psi \right)
-\frac{1}{2} \delta_V \partial_\nu \left(\bar{\psi} \gamma_\mu \psi \right)
 \partial^\nu \left(\bar{\psi} \gamma^\mu \psi \right)
  \nonumber \\
    &-&\frac{1}{2} \delta_{TV} \partial_\nu \left(\bar{\psi} \vec{\tau} \gamma_\mu \psi \right)
 \partial^\nu \left(\bar{\psi} \vec{\tau} \gamma_\mu \psi \right)
                                                      \nonumber\\
&-&\frac{1}{4} F^{\mu \nu} F_{\mu \nu} - e \frac{1-\tau_3}{2} \bar{\psi} \gamma^\mu \psi A_\mu ,
\label{Lag}\ee
where $M$ is the nucleon mass,
$\alpha_{S}$, $\alpha_{V}$ and $\alpha_{TV}$ represent the coupling constants
for four-fermion contact terms,
$\beta_{S}$, $\gamma_{S}$ and $\alpha_{V}$ are those for the higher-order terms
which are responsible for the effects of medium dependence,
and $\delta_{S}$, $\delta_{V}$ and $\delta_{TV}$ represent the gradient terms
which are included to simulate the finite range effects.
$A_\mu$ and $F_{\mu \nu}$ are respectively, the four-vector potential
and field strength tensor of the electromagnetic field.
The subscripts $S, \, V$ and $TV$ stand for scalar, vector and isovector, respectively.

By applying the mean-field theory to the Lagrangian density~(\ref{Lag}), 
the Hamiltonian density can be obtained by the Legendre transformation.
The RHB equation for the nucleons can be derived from the variational principle as follows~\cite{Kucharek1991}, 
\begin{gather}
\begin{pmatrix}
h_D - \lambda & \Delta \\ - \Delta^* & -h^*_D + \lambda
\end{pmatrix}
\begin{pmatrix}
U_k \\ V_k
\end{pmatrix}
= E_k \,
\begin{pmatrix}
U_k \\ V_k
\end{pmatrix}
\end{gather}
where $E_k$ is the quasiparticle energy,
$U_k$ and $V_k$ are the quasiparticle wave functions,
and $\lambda$ is the Fermi energy.
The Dirac Hamiltonian $h_D$ is
\be
h_D = \bm{\alpha} \cdot \bm{p} \, + \, \beta \left(M+S(\bm{r})\right) \, + \, V(\bm{r})
,
\ee
where the scalar and vector potentials are, respectively,
\begin{subequations} 
\be 
S(\bm{r}) &=& \alpha_S \rho_S \,+\, \beta_S \rho^2_S \,+\,
          \gamma_S \rho^3_S \,+\, \delta_S \Delta\rho_S, \label{sPot} \\ 
V(\bm{r}) &=& \alpha_V \rho_V \,+\, \gamma_V \rho^3_V \,+\, \delta_V \Delta \rho_V
          \,+\, e A_0 \,   \nonumber  \\   && +\, \alpha_{TV} \tau_3 \rho_{TV} \,+\, \delta_{TV} \tau_3 \Delta \rho_{TV}  . \label{vPot}
\ee
\end{subequations}
The local densities in Eqs.(\ref{sPot}) and (\ref{vPot}) are defined by 
\begin{subequations}
\be
\rho_S(\bm{r}) = \sum_{k>0} \, \bar{V_k}(\bm{r}) V_k(\bm{r}), \\
\rho_V(\bm{r}) = \sum_{k>0} \, V^\dag_k (\bm{r}) V_k(\bm{r}), \\
\rho_{TV}(\bm{r}) = \sum_{k>0} \, V^\dag_k (\bm{r}) \tau_3  V_k(\bm{r}),
\ee
\end{subequations}
which are constructed by the quasiparticle wave functions.

The pairing potential for particle-particle channel reads
\be
\Delta_{k k^{'}}(\bm{r},\bm{r^{'}}) = - \sum_{\tilde{k}\tilde{k}^{'}} \,
                                 \bm{V}_{k k^{'},\tilde{k}\tilde{k}^{'}} (\bm{r},\bm{r^{'}})
                                 \kappa_{\tilde{k}\tilde{k}^{'}} (\bm{r},\bm{r^{'}}),
\ee
which depends on the pairing tensor $\kappa=U^{\ast}V^T$~\cite{RingB1980}.
We use the density-dependent delta pairing force
\be
V^{pp} (\bm{r},\bm{r^{'}}) = \frac{V_0}{2} \left(1 - P^{\sigma} \right)
                             \delta (\bm{r} - \bm{r^{'}} )
                             \left(1 - \frac{\rho(\bm{r})}{\rho_{sat}} \right)
, 
\ee
where $\rho_{sat}$ is the nuclear matter saturation density. 
\textcolor{black}{
The total energy of a nucleus is calculated by~\cite{MengB2016,MengRHB1998} 
\begin{eqnarray}
E&=& \sum_{k>0}(\lambda_{\tau}\! - \! E_k)v_k^2 - E_{\mathrm{pair}} +E_{\mathrm{c.m.}} 
-\int \mathrm{d}^3\mathbf{r} \left[ \frac{1}{2}\alpha_S\rho^2_S \right. 
\nonumber \\   
 &+&
\frac{1}{2}\alpha_V\rho_V^2 + \frac{1}{2}\alpha_{TV}\rho_3^2 + \frac{2}{3}\beta_S\rho_S^3 +\frac{3}{4}\gamma_S\rho_S^4 +\frac{3}{4}\gamma_V\rho_V^4 
\nonumber \\   
 &+&  
  \! \left. \frac{1}{2} ( \delta_S\rho_S\Delta\rho_S \! 
+ \! \delta_V\rho_V\Delta\rho_V\! +\! \delta_{TV}\rho_3\Delta\rho_3\! +\! \rho_peA^0) \right]  
.
\end{eqnarray} 
The zero-range pairing force results in a local pairing field $\Delta(\mathbf{r})$. 
The pairing energy is given by 
\be 
E_{\mathrm{pair}} = -\frac{1}{2}\int\mathrm{d}^3\mathbf{r} \kappa(\mathbf{r})\Delta(\mathbf{r})
. 
\ee
The center-of-mass (c.m.) correction is calculated microscopically from the expectation value of the total momentum in the c.m. frame, 
\be 
E_{\mathrm{c.m.}}=-\langle {\hat{\mathbf{P}}}^2 \rangle / 2mA.
\ee 
}

Axial deformation with spatial reflection symmetry is taken into account
by expanding the  potential ($S(\bm{r})$, $V(\bm{r})$)
and densities ($\rho_S(\bm{r})$, $\rho_V(\bm{r})$, $\rho_{TV}(\bm{r})$)
in terms of the Legendre polynomials~\cite{Price1987},
\be
f(\bm{r}) = \sum_{\lambda} \, f_{\lambda} (r) P_{\lambda} (\rm{cos}\theta),
            \,\,\, \lambda = 0, \, 2, \, 4, \, \cdots.
\ee
For the description of nuclei close to the drip line it is important to include consistently
both the continuum and deformation effects
and the coupling among all these features.
A full treatment of the continuum by solving the equations in coordinate space, as done in the spherical case within the RCHB model~\cite{Meng1996,MengRHB1998}, is not feasible at present.
Instead, in the deformed case the continuum is taken into account by expanding the wavefunctions in the Dirac Wood-Saxon (WS) basis~\cite{Zhou2003}.
The present numerical implementation is therefore called DRHBc, standing for Deformed Relativistic Hartree-Bogoliubov theory in continuum~\cite{Li2012}.

\subsection{Numerical details} 
The numerical code \textcolor{black}{for applying DRHBc and various tests, including convergence tests with respect to the basis and the Legendre expansion, are presented in Ref.~\cite{Zhang2020}.
Following those tests, for convergence of the DRHBc calculations,
we use
the angular momentum cutoff $J_{\max}=23/2 \, \hbar$
and
the expansion order $\lambda_{\max}=6$.
The energy cutoff for the Dirac WS basis is taken as $E_{\mathrm{cut}}=300$ MeV.
At present we use the density functional PC-PK1~\cite{Zhao2010}
for the particle-hole channel.
For the particle-particle channel,
the zero-range pairing force with saturation density $\rho_{sat}=0.152 \,\, \rm{fm^{-3}}$ ~\cite{Xia2018}
and the strength $V_0 = -325 \,\, \rm{MeV}\,\rm{fm}^{\rm{3}}$
is used.
Since we use a zero-range pairing force,
we have to introduce a pairing cutoff over quasiparticle space.
Here,
a sharp cutoff is adopted with a pairing window of 100 MeV.
The pairing strength was determined so as to reproduce experimental even-odd mass differences within the above-defined model space~\cite{Zhang2020}.
} 

For each isotope, unconstrained calculations are performed by using different initialization conditions, in particular, different initial deformation parameter values 
$\beta = -0.4, -0.2,\ldots 0.6$ for the initial potential. 
These calculations may or may not lead to the same solution, depending on how many local minima exist in the isotope's energy landscape. 
In spherical cases (such as O and Ca isotopes) typically all solutions converge to the spherical ground state. 
In the case of more than one solution with very similar energies the ground state can be confirmed with constrained calculations, 
whereby the target deformation parameter is fixed. 
In this way, the dependence of the energy on the deformation parameter and the absolute minimum value of the energy can be verified. 
The presence of near-degenerate solutions is interesting in the context of shape coexistence.

We note that in the spherical RCHB applications of Ref.~\cite{Xia2018} the PC-PK1 density functional was used as well.
However, the pairing strength determined in the same way was $-342.5 \,\, \rm{MeV}\,\rm{fm}^{\rm{3}}$,
while the angular momentum cutoff was set to $19/2 \, \hbar$.
Thus comparisons between DRHBc and RCHB results will reflect the deformation and pairing effects combined. 
In order to test overall consistency, we verified that total binding energies of closed-shell nuclei from DRHBc and those from RCHB are the same (within tens of keV).


\section{Results and discussion\label{Sec:Results}}

\subsection{Deformation along isotopic chains} 

\begin{figure}
\centering
\resizebox{0.42\textwidth}{!}{%
  \includegraphics{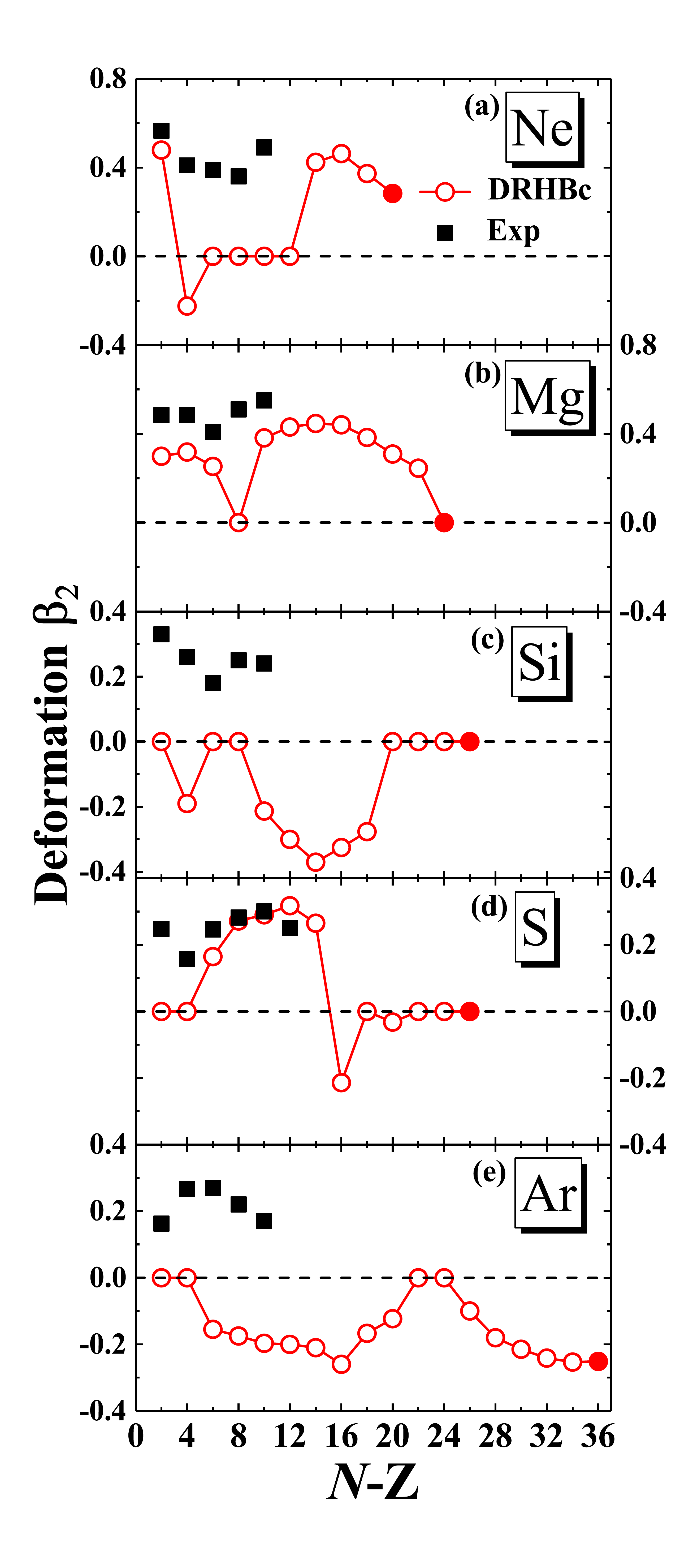}
}
\caption{
Deformation parameters $\beta_2$ 
as a function of the difference between the neutron and proton numbers  $N-Z$ 
for (a) Ne, (b) Mg, (c) Si, (d) S and (e) Ar isotopes. 
The experimental data are taken from the NNDC~\cite{NNDC}. 
Filled circles denote the first isotope beyond the drip line. 
\label{Fig-All-Beta}
}
\end{figure}

The main purpose of the present work is
to investigate the effect of the deformation degree of freedom on the neutron drip line location for even-even isotopes
from O to Ca. 
We begin by inspecting the evolution of deformation along the isotopic chains for $N \geq Z$. 
The O and Ca isotopes are predicted spherical by DRHBc and therefore no changes are expected with respect to RCHB predictions.
The quadrupole deformation parameters for the Ne, Mg, Si, S, and Ar isotopic chains are shown in Fig.~\ref{Fig-All-Beta} 
up to one isotope beyond the predicted drip-line nucleus. 
Many isotopes, including very neutron-rich ones, are predicted deformed. 
The proton and neutron deformation parameters ($\beta_{2,p}$ and $\beta_{2,n}$) are similar to each other (though not equal) in all the cases shown here,
and we show the weighted average deformation parameter given by $\beta_2=(Z/A)\beta_{2,p}+(N/A)\beta_{2,n}$.
All the experimental data are taken from the NNDC~\cite{NNDC}. 

We note that only the absolute value, not the sign, of the deformation parameter is determined experimentally. 
In addition, the definitions of the experimental and the calculated parameters are different: The experimental $\beta_2$ parameter is determined from the electric quadrupole excitation spectrum while the theoretical one quantifies the degree of the intrinsic deformation of the mean field. Therefore, the values do not necessarily agree~\cite{Raman2001}. 
Note that the experimental $\beta_2$ values of intrinsically spherical nuclei are not necessarily zero.

\begin{figure}
\centering
\resizebox{0.42\textwidth}{!}{%
  \includegraphics{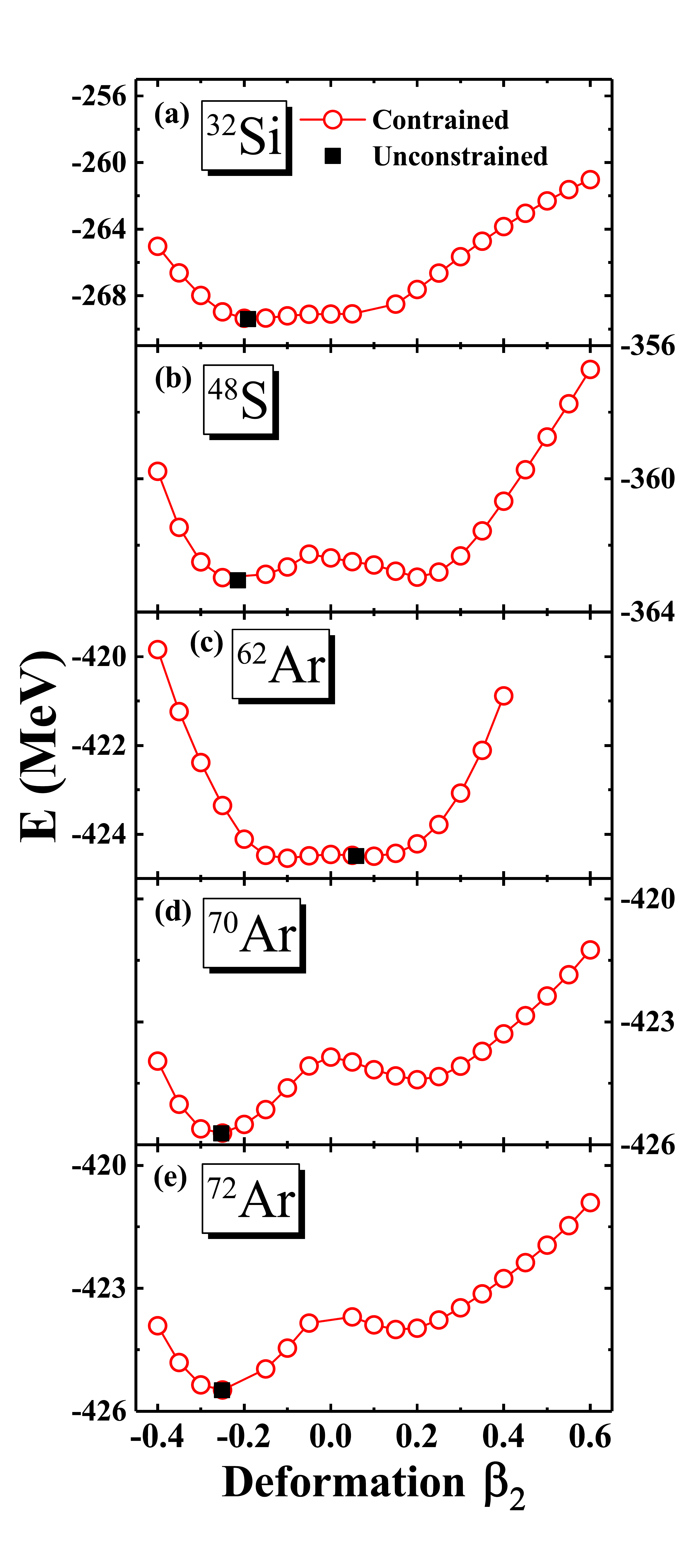}
}
\caption{
Energy as a function of the deformation parameter $\beta_2$  
obtained via constrained DRHBc calculations for selected isotopes: 
(a) $^{32}_{14}$Si$_{18}$, 
(b) $^{48}_{16}$S$_{32}$, 
(c) $^{62}_{18}$Ar$_{44}$, 
(d) $^{70}_{18}$Ar$_{52}$, 
and
(e) $^{72}_{18}$Ar$_{54}$ (unbound).
The solution obtained by unconstrained calculations is also indicated. 
\label{Fig-Constrained}
}
\end{figure}

In some cases the deformation parameter seems to oscillate at random between neighboring isotopes, for example, around $^{24}_{10}$Ne$_{14}$ or $^{48}_{16}$S$_{32}$. It is because we show only the deformation parameter of the very lowest-energy solution in each case. However, 
the $8<Z<20$ region examined here includes several candidates for shape coexistence or softness against deformation, which is borne out of our calculations. 
We find several isotopes with near-degenerate local energy minima (with energy difference of up to a few hundred keV) for different deformation parameters. 
(We stress that there is no ambiguity for the predicted location of the drip line as a result.) 
We show in Fig.~\ref{Fig-Constrained} the energies of some representative isotopes as a function of deformation parameter, obtained by constrained DRHBc calculations.  
For example, although $^{32}_{14}$Si$_{18}$ is indicated as oblately deformed in Fig.~\ref{Fig-All-Beta}(c), it is revealed that the energy of the spherically symmetric solution is within about 300~keV from the oblate solution. The energies of the local minima near $\beta_2 = \pm 0.1$ for $^{62}_{18}$Ar$_{44}$ differ by less than 100~keV. 
The examples shown in Fig.~\ref{Fig-Constrained} include very neutron-rich Ar isotopes for the discussion in Sec.~\ref{Sec:SepEn}. 

\subsection{Location of the drip line\label{Sec:dripline}} 

The theoretical prediction for the location of the neutron drip line is determined by inspecting
both the two neutron separation energy $S_{2n}$ and the neutron Fermi energy $\lambda_n$.
The former is defined as follows,
\be
S_{2n}(Z,N) \,=\, B(Z,N)-B(Z,N-2),
\ee
where \textcolor{black}{$B(Z,N)=-E(Z,N)$} is the binding energy of the nucleus with atomic number $Z$ and neutron number $N$. 
When the two neutron separation energy is found positive
and the neutron Fermi energy is found negative,
the nucleus is classified as bound.
The last bound isotope found along an isotopic chain defines the location of the neutron drip line. 
The heaviest even isotopes of the O--Ca  elements discovered so far are
$_8^{26}$O$_{18}$ (somewhat unbound),
$^{34}_{10}$Ne$_{24}$,
$^{40}_{12}$Mg$_{28}$,
$^{44}_{14}$Si$_{30}$,
$^{48}_{16}$S$_{32}$,
$^{54}_{18}$Ar$_{36}$,
and
$^{60}_{20}$Ca$_{40}$~\cite{Lunderberg2012,Notani2002,Ahn2019,Baumann2007,Tarasov2007,Lewitowicz1990,Tarasov2018,Thoen2012a,Thoen2012b}.

\textcolor{black}{In Ref.~\cite{Xia2018} it was shown that the coupling of bound states with continuum states 
due to pairing correlations can extend considerably the theoretically predicted nuclear landscape. 
In Ref.~\cite{Qu2013}, 
which focused on the region from Oxygen to Titanium, 
RCHB results were compared with those of various mass models with no explicit treatment of the continuum. 
It was concluded that including a proper description of the continuum can extend the neutron drip line by several isotopes for each element.  
}

Here we explore the deformation effects on the neutron drip line location
by comparing the results from the DRHBc theory
with those from the RCHB theory~\cite{Xia2018}. The predicted location of the drip line will of course depend on the density functional parameterization used (here, PC-PK1).
In this work, we are mainly interested in possible deformation effects on the neutron drip lines in a relative sense.
\begin{figure}
\centering
  \includegraphics[width=0.43\textwidth]{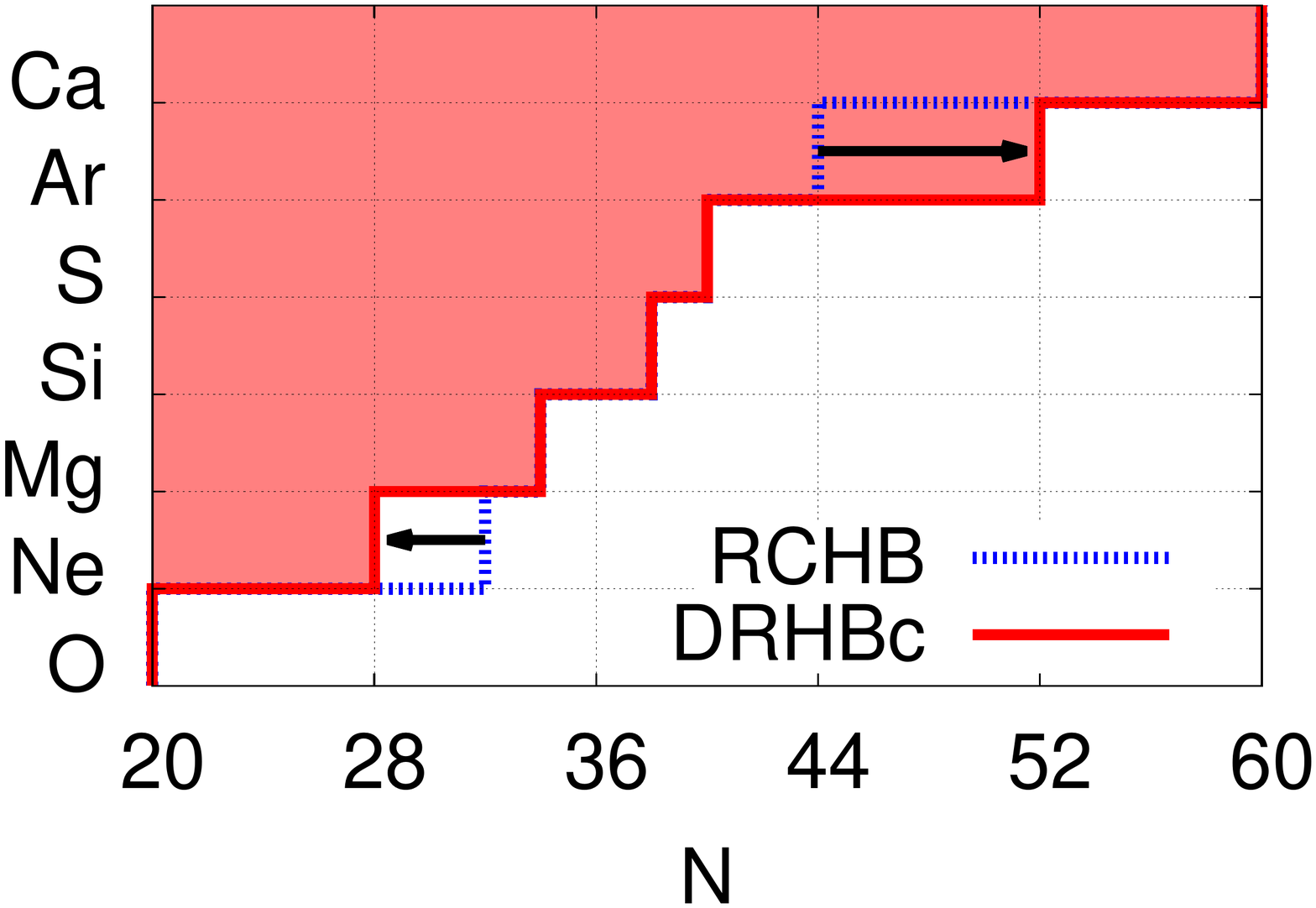} \\[-2mm] 
\mbox{~} \,\, 
\begin{tabular}{|c|c|c|}
\hline
\multirow{2}{*}{Element (Z)} & \multicolumn{2}{|c|}{neutron number (N)} \\
\cline{2-3}
{} & RCHB & DRHBc \\
\hline
O (8) & 20 & 20  \\
\hline
Ne (10) & 32 & 28 \\
\hline
Mg (12) & 34 & 34 \\
\hline
Si (14) & 38 & 38  \\
\hline
S (16) & 40 & 40 \\
\hline
Ar (18) & 44 & 52 \\
\hline
Ca (20) & 60 & 60 \\
\hline
\end{tabular}
\caption{
The neutron drip-line location
predicted in the present work by the DRHBc theory (with deformation)
in comparison with the RCHB results (without deformation)~\cite{Xia2018}
for even-even isotopes with $8 \leq Z \leq 20$. 
The arrows show the direction of the changes
with the inclusion of deformation. 
The neutron numbers at the drip line are tabulated below the figure. 
\label{Fig-drip}
}
\end{figure}

The results are summarized and visualized in Fig.~\ref{Fig-drip}.
In all the cases the predicted drip line lies beyond the isotopes experimentally discovered so far.
The drip-line locations are not affected with the inclusion of deformation,
except for the Ne and Ar isotopic chains.
As expected, the drip O and Ca isotopes are predicted the same in both cases, because all those isotopes are predicted spherical.
We also find that the last two or three bound S and Si isotopes are predicted spherical (see Fig.~\ref{Fig-All-Beta}).
On the contrary, nuclei in the vicinity of the drip line are predicted deformed for Ne, Mg, and Ar.

It is especially interesting that, 
while the Ar isotopic chain is extended to higher $N$ when deformation is included, 
\textcolor{black}{augmenting the continuum effects},
the neutron drip line shrinks in the case of Ne isotopes.
There is no change in Mg due to the inclusion of deformation.
We therefore proceed to inspect these isotopes in more detail.

\subsection{Evolution of separation energies\label{Sec:SepEn}: Ne, Mg, and Ar}

\begin{figure}
\centering
\resizebox{0.45\textwidth}{!}{%
  \includegraphics{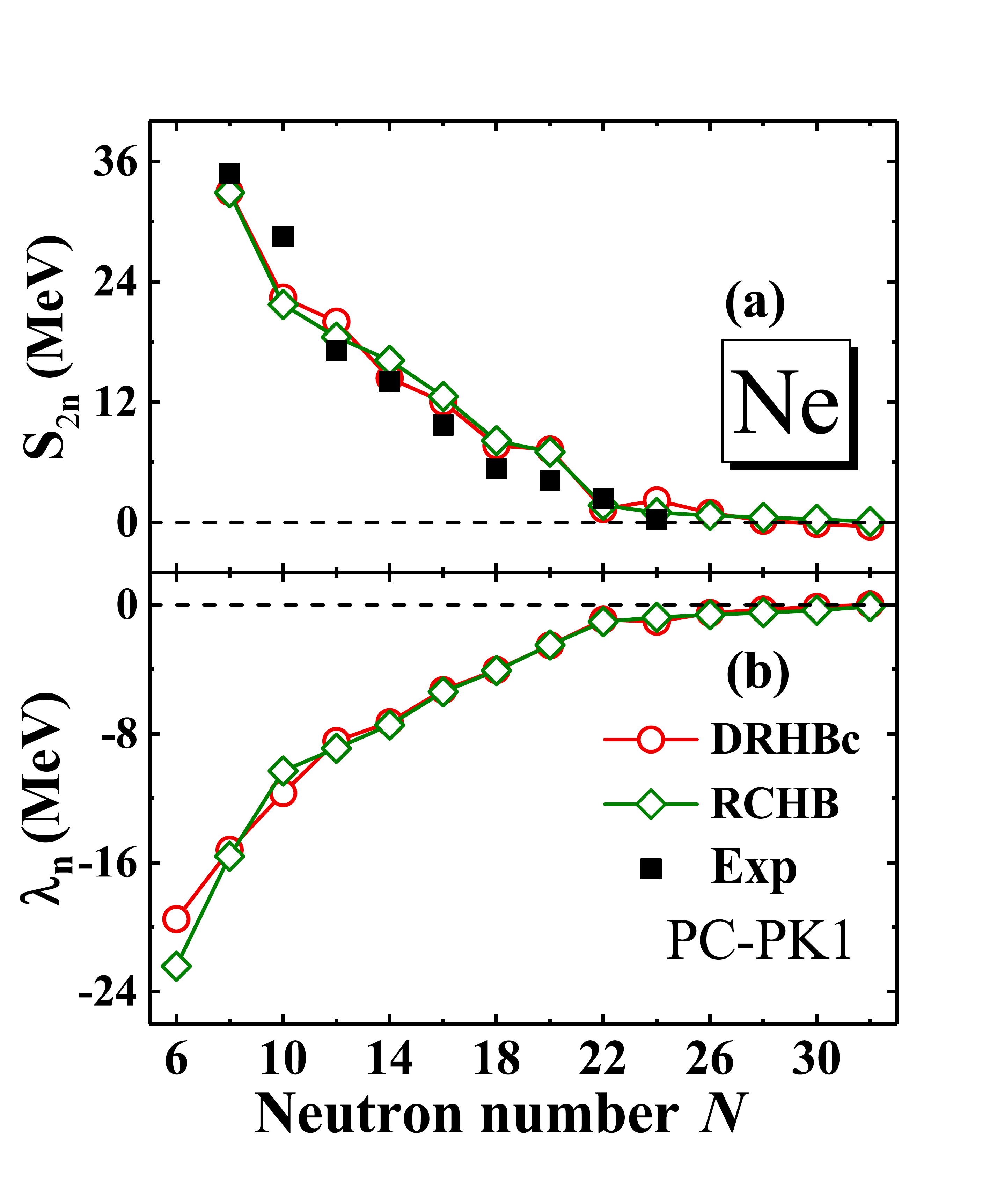}
}
\caption{
(a) Two neutron separation energy ($S_{2n}$)
and (b) neutron Fermi energy ($\lambda_{n}$) of Ne isotopes.
The red lines with empty circles are for DRHBc calculations and
the green lines with empty diamonds are for RCHB calculations.
The experimental data taken from the NNDC~\cite{NNDC} are shown by the black filled squares.
\label{Fig-Ne-S2nL}
}
\end{figure}

For Ne, Mg, and Ar isotopes we inspect the quantities $S_{2n}$ and $\lambda_n$ which define the location of the drip line.
Figure~\ref{Fig-Ne-S2nL} (a) shows
two neutron separation energy ($S_{2n}$) of Ne isotopes
with respect to the neutron number.
In the RCHB calculations (green lines),
two neutron separation energies ($S_{2n}$)
remain positive until the neutron number 32.
However,
the value of $S_{2n}$ from the DRHBc calculations (red lines) remain positive up to the neutron number 28.
Figure~\ref{Fig-Ne-S2nL} (b) shows
neutron Fermi energies ($\lambda_n$) of Ne isotopes and
they stay negative up to the neutron number $32$ in the RCHB calculation.
However, they stay negative up to $30$ in the DRHBc calculations.
Therefore, the neutron drip-line location
of Ne isotopes
is predicted to be $^{38}$Ne instead of $^{42}$Ne.

\begin{figure}
\centering
\resizebox{0.45\textwidth}{!}{%
  \includegraphics{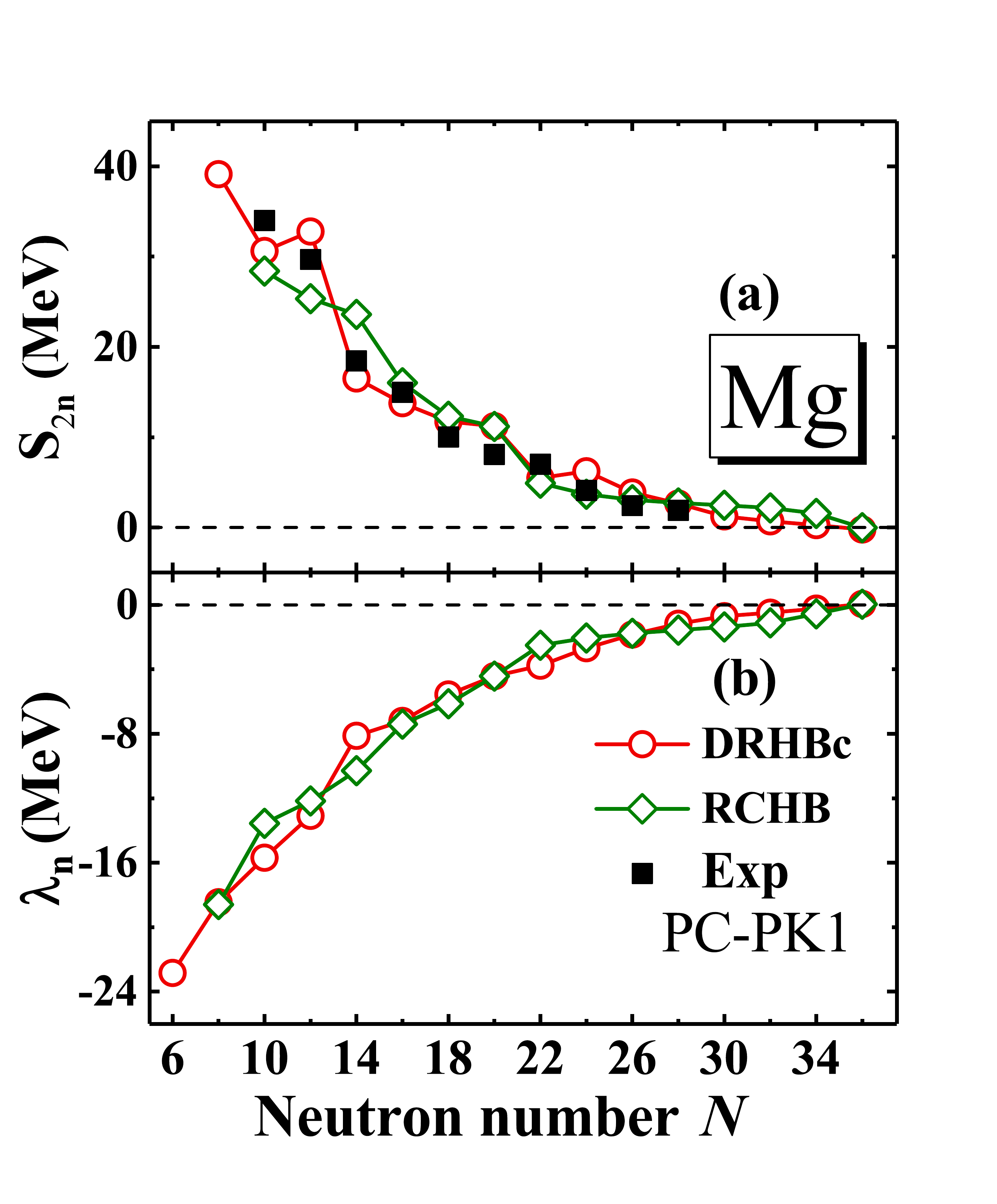}
}
\caption{
Same as Fig.~\ref{Fig-Ne-S2nL} but for Mg.
\label{Fig-Mg-S2nL}
}
\end{figure}

Figure~\ref{Fig-Mg-S2nL} shows the same quantities for Mg isotopes. There are sizable differences in the predictions towards the drip line but not for the drip nucleus, $N=34$. In addition, the next isotope, unbound with $N=36$, is predicted spherical. Therefore, the situation turns out to be similar to the Si, S cases.

\begin{figure}
\centering
\resizebox{0.45\textwidth}{!}{%
  \includegraphics{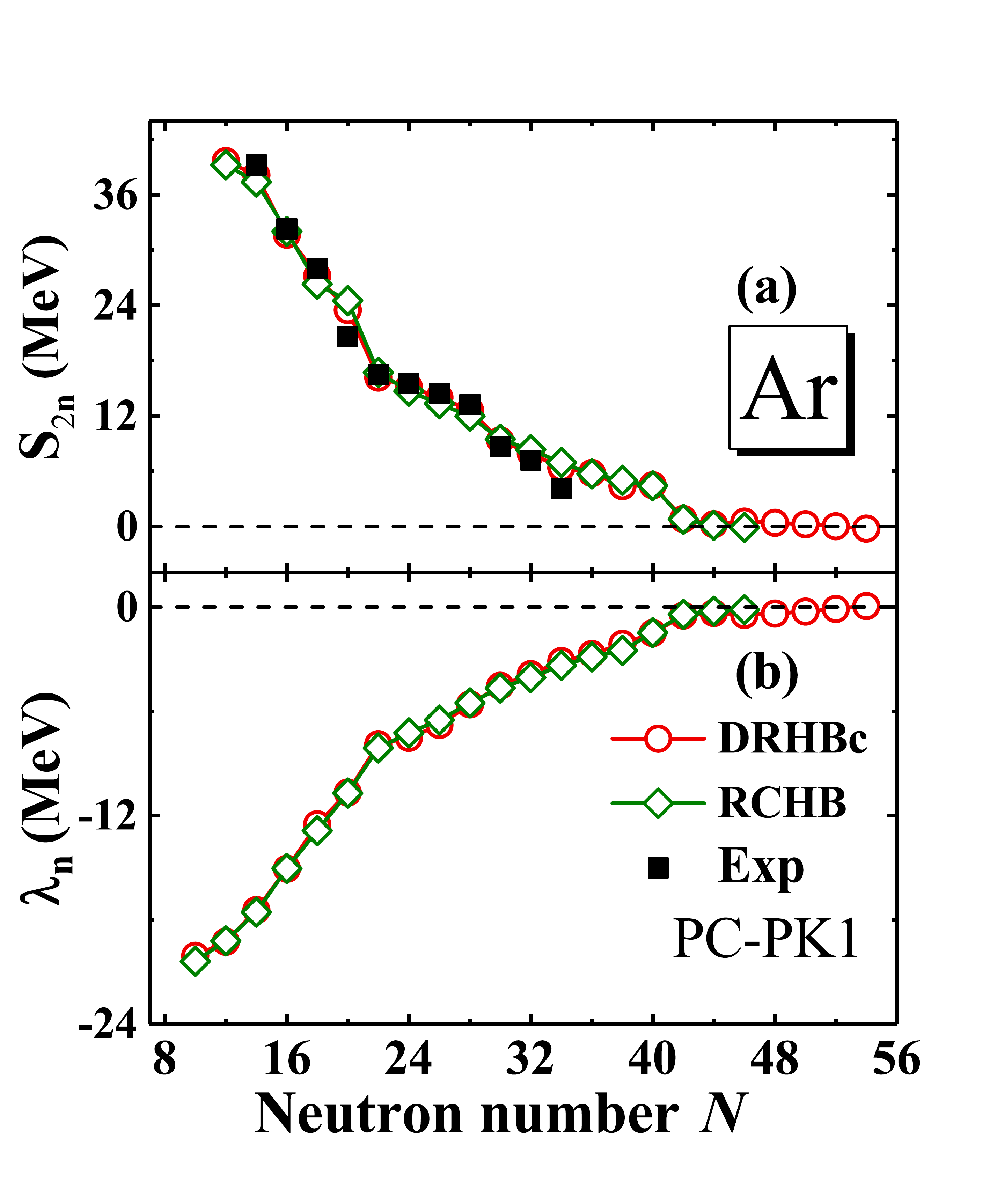}
}
\caption{
Same as Fig.~\ref{Fig-Ne-S2nL} but for Ar.
\label{Fig-Ar-S2nL}
}
\end{figure}

Figure~\ref{Fig-Ar-S2nL} shows the same quantities for Ar isotopes.
In the RCHB calculations $S_{2n}$
remains positive until the neutron number becomes 44, while in the DRHBc calculations $S_{2n}$ is positive until the neutron number 52.
Similarly, $\lambda_n$ values of Ar isotopes
stay negative up to the neutron number $46$ in the RCHB calculations and up to $52$ in the DRHBc calculations.
Therefore, the neutron drip-line location
of Ar isotopes
is predicted to be $^{70}$Ar instead of $^{62}$Ar.

For both Ne and Ar the differences between the two models appear marginal at large $N$, but they are sufficient to shift the drip-line location.
Furthermore, the differences in the proximity of the drip line have opposite signs for Ne and Ar leading to the shrinking of the drip line in the former case when deformation is taken into account and to extension in the latter. The question arises as to whether such results are consistent with the generally stronger binding expected from the DRHBc calculations.

Indeed, when deformation is included in the calculations, but everything else (interaction, cut-off parameters, etc.) kept the same as in the calculations in spherical symmetry and if a deformed solution is found to be the ground state, the deformed state must be more bound than the spherical solution by definition.
To put it differently, by including more degrees of freedom (deformation) we expect equal or lower energies by virtue of the variational principle.
Then generally the DRHBc calculations should give more (or equally) bound solutions than the RCHB calculations.
(Exceptions may still occur, especially for near-spherical nuclei, owing to the different numerical implementations and different pairing parameters used in the two types of calculations.)
As s result, an extension of the drip line might be expected rather than a shrinking.
On the other hand, it is not the value of the binding energy that matters, but the difference between the binding energies of neighboring isotopes.
The difference between the two-neutron separation energy predicted by DRHBc,
$S_{2n}$(DRHBc), and that predicted by the spherical model RCHB, $S_{2n}$(RCHB),
is given by
\begin{eqnarray}
\Delta S_{2n} &\equiv &  S_{2n}{\mathrm{(DRHBc)}} - S_{2n}\mathrm{(RCHB)} \nonumber \\
                       &  =   &  \Delta B(Z,N) - \Delta B(Z,N-2), 
\label{Eq:DSDB} 
\end{eqnarray}
where we have denoted by $\Delta B = B$(DRHBc)$-B$(RCHB) the difference in the binding energies of each model.
Even though both $\Delta B(Z,N)$ and $\Delta B(Z,N-2)$ are expected to be positive, the difference $\Delta S_{2n}$ can be either positive or negative.
If $\Delta B(Z,N)$ is smaller than $\Delta B(Z,N-2)$,
then the drip line may shift to lower $N$.
Let us therefore investigate $\Delta B$ and $\Delta S_{2n}$.

\begin{figure}
\centering
\resizebox{0.45\textwidth}{!}{%
  \includegraphics{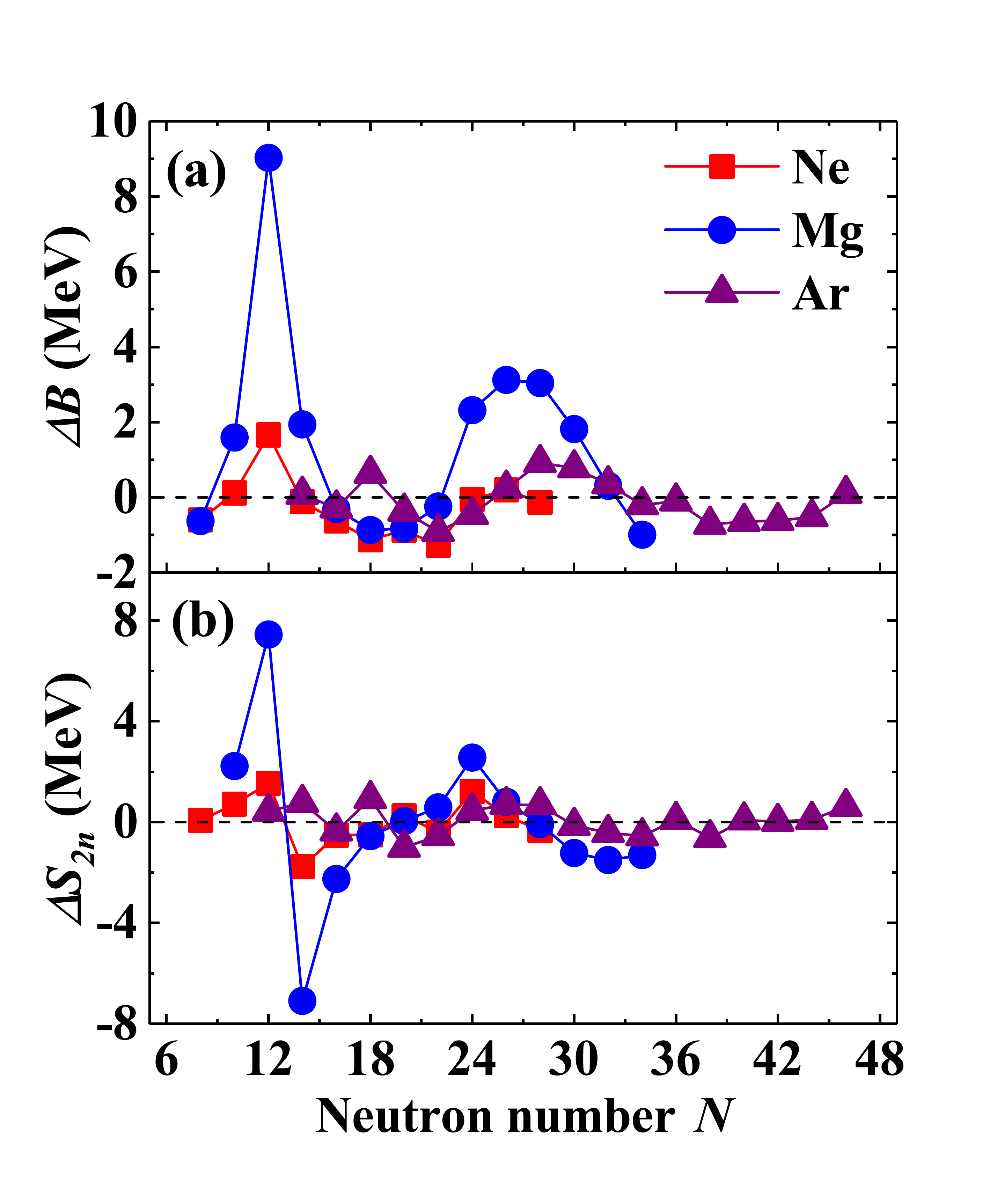}
}
\caption{
\textcolor{black}{The differences $\Delta B$ and $\Delta S_{2n}$ between results of the DRHBc calculations (with $V_0=-325\,\mathrm{MeV\, fm^3}$) and the RCHB calculations (with $V_0 =-342.5\,\mathrm{MeV\, fm^3}$).} 
\label{Fig-All-BE}
}
\end{figure}

The difference in binding and separation energies for the three isotopic chains Ne, Mg, and Ar is shown in Fig.~\ref{Fig-All-BE}. 
\textcolor{black}{
As expected from Eq.~(\ref{Eq:DSDB}) $\Delta S$ resembles a differential of $\Delta B$: an increase (decrease) of $\Delta B$ with $N$ leads to a positive (negative) value for $\Delta S$. 
} 
For Ar, the separation energy difference increases towards the drip line.
Thus, for example, $\Delta B(Z=18,\,N=46)$ is larger than $\Delta B(Z=18,\,N=44)$.
The increase in the gain of binding owing to deformation is consistent with the increase in the (absolute value of the)
deformation parameter. (See Fig.~\ref{Fig-All-Beta}(e).) 
The evolution of the Ar energy landscape towards and beyond the neutron drip line is also exemplified by Fig.~\ref{Fig-Constrained}(c),(d), and (e).
$^{62}$Ar represents a transitional situation between the spherical $^{58,60}$Ar isotopes and the heavier oblate isotopes, whose 
absolute deformation paramater increases with $N$ until $N$ becomes equal to $52$ and the drip line is reached.
Such a trend can lead to an extension of the drip line.
On the contrary, for Ne the trend of the separation energy difference is to drop towards the drip line.
Thus $\Delta B(Z=10,\,N=28)$ is smaller than $\Delta B(Z=10,\,N=26)$.
The drop in the gain in binding owing to deformation is consistent with the decrease in the
deformation parameter. (See Fig.~\ref{Fig-All-Beta}(a).)
Such a trend can lead to the shrinking of the drip line. 
As obvious from the Mg case, the above are not sufficient conditions for the respective results to occur.
However, the condition  $ \Delta S_{2n} <0$ seems necessary for the shrinking of the drip line.

At this point a few comments are necessary on what appear to be possible violations of the variational principle. 
Fig.~\ref{Fig-All-BE}(a) 
\textcolor{black}{shows that $\Delta B$ becomes negative for}   
some neutron numbers,  in which case the DRHBc solutions are less bound than the RCHB ones. First, as already mentioned, some minor numerical discrepancies are expected from the model space cut-off parameters whose optimal values differ for each of the numerical implementations and are related to the WS basis treatment of continuum states.
More importantly, the optimal pairing parameter for DRHBc is found weaker than in RCHB leading to less binding coming from pairing correlations.
In order to investigate this effect, 
\textcolor{black}{
we have performed the same calculations and comparisons by using the same pairing strength as in RCHB, $V_0 =-342.5\,\mathrm{MeV\, fm^3}$. 
We found that the general trends for $\Delta B$, $\Delta S_{2n}$ do not change. 
The main difference is a small positive shift for $\Delta B$, leading to better compliance with the variational principle. 
}

\section{Conclusions\label{Sec:Conclusions}}
Even-even neutron-rich isotopes 
from O to Ca
were investigated by using the DRHBc theory with the PC-PK1 functional. 
The neutron drip-line location was determined by
calculating the two-neutron separation energies and
the neutron Fermi energies.
In order to investigate the deformation effect, 
on the neutron drip-line location,
we compared the present results with those predicted by the RCHB theory with spherical symmetry.
We found that the Ne and Ar drip line nuclei are different when the deformation effect is included. 
The direction of the change in the neutron drip line 
is not necessarily towards an extended drip line. 
It  rather appears dependent upon the evolution of the degree of deformation (magnitude of the deformation parameter) towards the drip line:
When the drip line nuclei are predicted spherical, the drip line doesn't change;  
for Ne the deformation decreases towards the drip line and the drip line ``shrinks"; the opposite is seen for Ar. 
We conclude that taking into account deformation effects as well as pairing and continuum effects in a consistent way can affect critically the theoretical description of the neutron drip-line location.

It would be interesting to see if similar trends are observed in different regions of the nuclear chart. 
Also shape coexistence and isomerism remain important topics for future work. 
Several investigations are currently in progress within the broader DRHBc mass table collaboration~\cite{Zhang2020,Sun2018,Zhang2019,Pan2019}.

\section*{Acknowledgments}

We thank the members of the DRHBc Mass Table Collaboration for useful discussions.
The work was supported partly by the Rare Isotope
Science Project of Institute for Basic Science funded by
Ministry of Science, ICT and Future Planning, and NRF of
Korea
(2013M7A1A1075764).
EJI and SWH were supported in part by the Korea government MSIT through the
National Research Foundation (2018M2A8A2083829).

\end{document}